\documentclass[aps,prb,twocolumn,notitlepage,longbibliography,showpacs,superscriptaddress,floatfix]{revtex4-1}

\usepackage{graphicx}
\usepackage{amsmath,amssymb,bbold,bm,color}
\usepackage{float}
\usepackage{epstopdf}
\usepackage{hyperref}
\usepackage{color}
\usepackage{enumerate}
\usepackage{wasysym}
\usepackage{url}
\usepackage{dsfont}
\usepackage{braket}
\usepackage[caption=false, justification=raggedright]{subfig}

\newcommand{\br}{\bf r}

\hypersetup{colorlinks=true,linkcolor=blue,citecolor=blue,urlcolor=blue}

\begin{document}

\title{Majorana zero modes in a magnetic and superconducting hybrid vortex}

\author{Vedangi Pathak}
\affiliation{Department of Physics and Astronomy \& Stewart Blusson Quantum Matter Institute, University of British Columbia, Vancouver BC, Canada V6T 1Z4}
\author{Sayak Dasgupta}
\affiliation{Department of Physics and Astronomy \& Stewart Blusson Quantum Matter Institute, University of British Columbia, Vancouver BC, Canada V6T 1Z4}
\affiliation{Institute for Solid State Physics, University of Tokyo, Kashiwa 277-8581, Japan}
\author{Marcel Franz}
\affiliation{Department of Physics and Astronomy \& Stewart Blusson Quantum Matter Institute, University of British Columbia, Vancouver BC, Canada V6T 1Z4}

\date{\today}

\begin{abstract}
We propose and investigate a new platform for the realization of Majorana zero modes in a thin-film heterostructure composed of an easy-plane ferromagnet and a superconductor with spin-orbit coupling. The system can support an energetically favorable bound state comprising a magnetic and a superconducting vortex. We show that a hybrid vortex thus created can host a robust zero-energy Majorana bound state at its core over a wide range of parameters, with its partner zero mode located at the outer boundary of a disk-shaped topological region. We identify a novel mechanism underlying the formation of the topological phase that, remarkably, relies on the orbital effect of the magnetization field and not on the usual Zeeman effect. The in-plane components of magnetization couple to electrons as a gauge potential with non-zero curl, thus creating an emergent magnetic field responsible for the gapped topologically non-trivial region surrounding the vortex core. Our construction allows the mobility of magnetic vortices to be imposed on the Majorana zero mode at the core of the superconducting vortex. In addition, the system shows a rich interplay between magnetism and superconductivity which might aid in developing future devices and technologies.
\end{abstract}
\maketitle

\section{Introduction}

One of the primary thrusts in current condensed matter physics is the search for new platforms capable of supporting exotic quasi-particles of which the Majorana quasi-particle is of special significance. These emergent Majorana excitations are topologically protected against local perturbations and exhibit non-Abelian exchange statistics\cite{Alicea_2012,beenakker13,franz-review}, making them ideal to store and manipulate quantum information via topological qubits\cite{Stern2008}. 

Initial theoretical proposals\cite{Kitaev2001} showed that a spinless $p-$wave superconductor in one dimension and correspondingly a spinless $p_{x} +ip_{y}$ in two dimensions will host Majorana zero modes. A more realistic implementation of topological $p-$wave superconductivity is in the form of superconducting heterostructures. Such heterostructures require the presence of a large superconducting gap, strong spin-orbit coupling (SOC) and a time-reversal breaking field to obtain an effective topological superconductor\cite{lutchyn-sau-sds-mzm-wire,yuval-refael-oppen-mzm-wire}. 

While a simple construction, it was soon realized that it was hard to obtain semiconductor wires with large enough SOC. This prompted alternate proposals where the need for an SOC was circumvented by exchange coupling the spins in the superconductor to non-collinear magnetic islands. In particular, Choy $et~al$ \cite{beenakker-adatom} showed that in presence of a chain of magnetic adatoms, where the spins of the adatoms were not aligned, the s-wave order parameter converts to an effective spin polarized $p-$wave superconductor with Majorana modes at the end. This proposal was followed by a series of works where a similar $s-$wave to $p-$wave conversion was showed for a wire with a helical magnetic order, proximitized to an $s-$wave superconductor\cite{loss-helical,franz-helical,braunecker-helical}. In these constructions the helical magnetic state is self stabilized through an effective RKKY type interaction mediated by the superconductor. An analogous construction was used by Nakosai $et~al$ \cite{nagaosa2013} to obtain a chiral $p-$wave superconductor in two dimensions by proximitizing an $s-$wave superconducting slab to a magnetic slab with a non-coplanar Skyrmion like arrangement of magnetic moments. In this chiral $p-$wave state, a vortex core was predicted to host a Majorana Zero mode (MZM). Recent experiments have shown signatures of chiral topological phases in ferromagnetic islands \cite{Menard2017,Menard2019}.

Building on these efforts Yang $et~al$ \cite{yang-dws} constructed a heterostructure with a double winding skyrmion proximitized to an s-wave superconductor, which produced stable MZMs bound to the center of the Skyrmion core. The isolation of the MZM from the other states in the vortex core was enhanced by increasing the winding. However, it is difficult to generate Skyrmion textures with high winding numbers; if one tries to construct them by merging individual Skyrmions with a single winding, the Skyrmions tend to energetically favour annihilation. What works is a heterostructure comprising a single winding Skyrmion bound to a vortex in the s-wave superconductor. This construction hosts an MZM at the vortex core and the Skyrmion-vortex system is bound together through an exchange coupling  \cite{rex-gornyi-sws-vortex}. 

The nature of the coupling between the magnetic texture and the proximitized superconductor was studied in \cite{schaffer-magnetoelectric-sc}, where the authors showed that the magnetic moments generate supercurrents which in turn generate the interaction between the superconducting vortex and the underlying magnetic texture. The current-current interaction reduces to an effective magnetoelectric potential with a strength proportional to the SOC and size of the magnetic moments. Hals $et~al$ \cite{hals-skyrmion-vortex} used this term to produce an energetically stable bound state between a superconducting vortex and a magnetic Skyrmion, and further showed how this composite object can be manipulated by using a spin current to move the magnetic Skyrmion in the magnet. The vortex being bound to the Skyrmion core is dragged along \cite{hals-skyrmion-vortex}. These works proposed a setup where one could, at least theoretically, isolate an MZM and move it around freely.

\begin{figure}
    \centering
    \includegraphics[width=\linewidth]{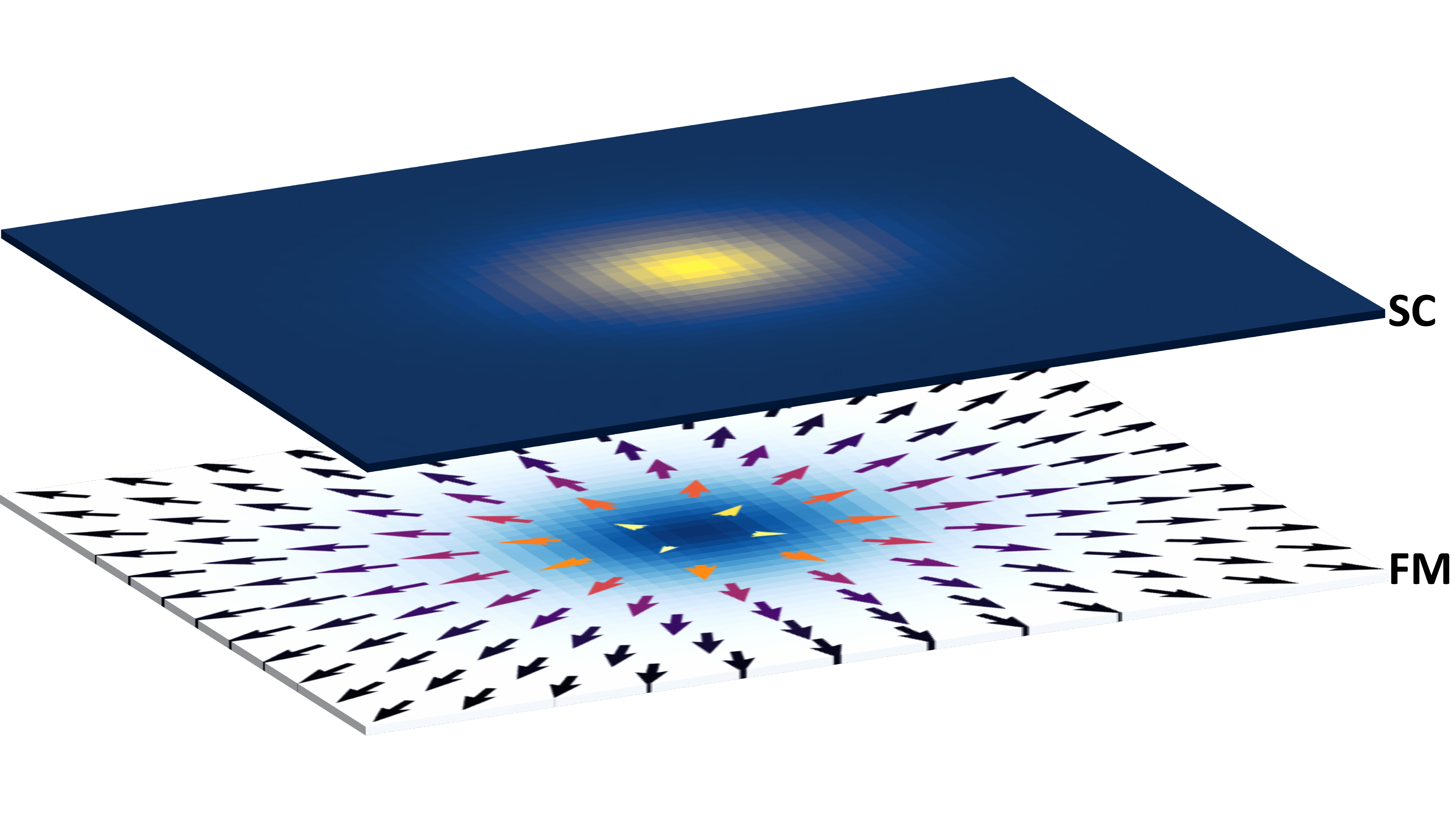}
    \caption{System comprising a ferromagnet-superconductor heterostructure. The magnetic and superconducting vortex form an energetically favourable hybrid vortex which hosts Majorana modes.}
    \label{fig:fig1}
\end{figure}

Inspired by this, here we investigate conditions for the stability of a hybrid magnetic and superconducting vortex pair in a heterostructure composed of ferromagnetic and superconducting thin films shown schematically in Fig.\ \ref{fig:fig1}. The magnetic vortex is a defect native to an easy plane ferromagnet. Such systems are rare in the natural state but recent studies have shown that monolayer van der Waals  materials such as NiPS$_3$ and CrCl$_3$ have strong easy plane nature, and nearly isotropic behavior with the easy plane \cite{XY-crcl3,XY-Van-der-Waals,Lu2020,mcguire2017}.

Another route to magnetic vortices is through domain walls in two dimensional uniaxial ferromagnets. For an easy axis in the $xy$ plane we have domain walls which interpolate between two states via a magnetic vortex and half vortices \cite{tretiakov2008}. These strips can be fairly wide (around $600$ nm) which is an order of magnitude larger than the typical magnetic vortex radius ($\sim 10$ nm) \cite{Beach2005}. Thus we can isolate the vortex core in the center of the magnetic strip which then interacts with the superconducting vortex. A similar domain wall hosting a vortex-like spin structure can form in the triangular chiral antiferromagnets Mn$_3$X (X=Sn/Ge). There the local easy axis anisotropy creates a D$_{3h}$ symmetric environment, producing six magnetic domains which meet at a vortex \cite{Liu:2017,Li2019}. At the vortex core the magnetic normal mode structure ensures a spin canting out of the kagome plane which provides the necessary Zeeman field to the superconducting vortex \cite{dasgupta-mn3ge-2020}. The domain wall version of the magnetic vortex is easier to generate than an isolated magnetic skyrmion, typically generated by melting Skyrmion crystals which exist in a very small temperature window of the phase diagram in a chiral magnet\cite{S.2009}. 

We want to couple the magnetic vortex to a superconducting vortex in an s-wave superconducting film. The superconducting gap of the order of $0.3-0.5$ meV \cite{Menard2019} is much smaller than the exchange interaction energies of the magnetic materials (particularly those where the vortex occurs at a domain wall) which are of the order of $10$ meV \cite{Herve2018}. So the magnetism is largely unaffected by the superconducting order parameter. In addition, the superconductor-ferromagnet heterostrucure needs to have a Rashba interaction in order to form a bound state \cite{schaffer-magnetoelectric-sc}. For an insulating ferromagnet this Rashba interaction can be sourced from the superconductor \cite{2dsc+rashba} or through an interfacial Dzyaloshinsky-Moriya interaction in the magnet. We use the former in our tight binding model.

We find that the two cores comprising the hybrid vortex can be strongly attractive, with an energy density that varies as a square of the core separation. The question then is if such a system can host a Majorana bound state, without an external Zeeman field, localized at the hybrid vortex, which we find to be the case. Specifically one zero mode is located at the core of the superconducting vortex and its partner at the outer ring of the topologically nontrivial region centered at the core. We find that the topological phase inside the ring is enabled by an emergent orbital magnetic field effect and, interestingly, not by the Zeeman field as would be the case in conventional models\cite{lutchyn-sau-sds-mzm-wire,yuval-refael-oppen-mzm-wire}.  Majorana bound states in such hybrid vortices are distinguished by their  ease of manipulation via the magnetic textures and by scalability. 

General conditions for stability of such a hybrid vortex are discussed in Section \ref{sec:free_energy}. Using a lattice model for this system (Section \ref{sec:model}), we then investigate the presence of Majorana bound states in a stable hybrid magnetic and superconducting vortex pair (Section \ref{sec:mzm}). 

\section{Hybrid vortex}
We consider a heterostructure made of an in-plane ferromagnet proximitized to an $s-$wave superconducting film. The ferromagnetic order parameter is represented by the magnetization field $\mathbf{m}(\mathbf{r})$ that is related to the spin field through the gyromagnetic ratio $\mathbf{m} = \gamma\mathbf{s}$. The magnetic soliton we want to couple to the superconductor is the vortex shown in Fig.\ \ref{fig:fig1}. The corresponding magnetization configuration, detailed below, is a solution to the Heisenberg energy density functional in two spatial dimensions \cite{dasgupta-2020-vortex} and forms naturally without the need for Dzylaoshinsky-Moriya interactions. 

The magnetic vortex has spins mostly in the $xy$ plane, except at the core of the vortex where the spins are forced to cant out of the plane by the exchange interaction. It is conveniently parameterized by two scalar fields
\begin{equation}\label{eq:spin_texture}
    \mathbf{m(r)}=m_{0}[\textrm{cos}\Phi\mathbf{(r)}\textrm{sin}\Theta\mathbf{(r)},\textrm{sin}\Phi\mathbf{(r)}\textrm{sin}\Theta\mathbf{(r)},\textrm{cos}\Theta\mathbf{(r)}].
\end{equation}
For the vortex centered at the origin the $\Theta(\mathbf{r})$ field obeys the boundary conditions $\Theta(\mathbf{r}\to \infty) = \pi/2$, and $\Theta(\mathbf{r} = 0) = 0$ or $\pi$. The $\Phi(\mathbf{r})$ field encodes the winding number defined as $ 2\pi n_m = \oint_c d\mathbf{r}\cdot \bm{\nabla}\Phi$, where the integration contour is taken to enclose the vortex center. The winding number can be any integer, but we restrict ourselves to  a singly quantized vortex or antivortex corresponding to $n_m = \pm 1$.

\begin{figure*}[hbt]
    \centering
    \includegraphics[width=\linewidth]{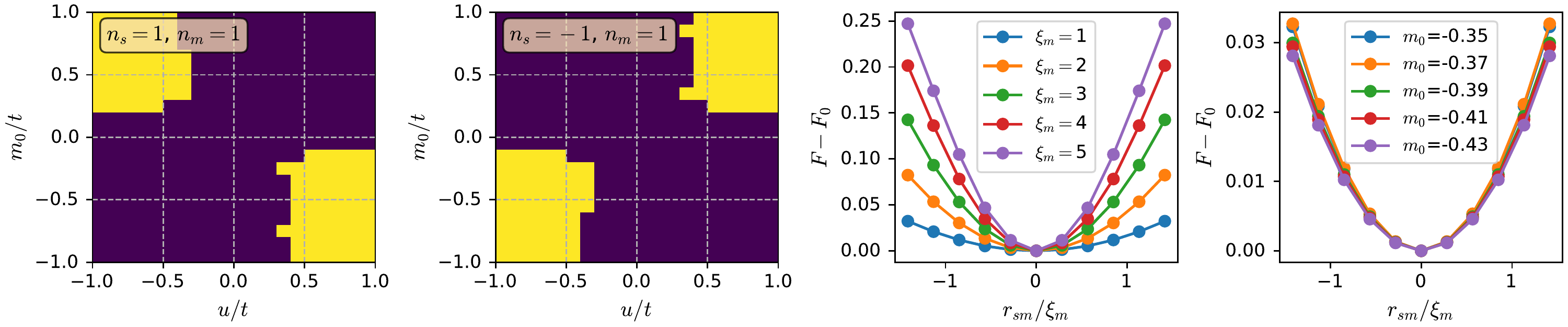}
    
    \caption{Left two panels: Energetically favorable regions in parameter space for hybrid vortex formation. With the magnetic vortex texture ($\xi_m=1$) fixed at the center of the system of size $L=21$, the superconducting vortex core ($\xi_s=1$) is moved along the diagonal starting at the position. The yellow shaded regions in all plots correspond to the points in the parameter space where the free energy minima roughly corresponds to the limit where the superconducting and the magnetic vortex are perfectly superimposed on each other (i.e. $r_{sm}=0$) thereby forming a stable hybrid vortex. Regions of stability calculated for a magnetic vortex and superconducting vortex pair ($n_s=n_m=1$) and for a magnetic vortex and a superconducting anti-vortex pair ($n_s=-1,n_m=1$). Right two panels: Calculation of free energy of the hybrid vortex as a function of the distance between the magnetic and superconducting vortex. Here, $F_0$ is the free energy when $r_{sm}=0$. With the magnetic vortex texture fixed at the center of the system, the superconducting vortex core is moved along the diagonal. Free energy is plotted for different magnetic decay length parameters $\xi_m$ at fixed $m_0=-0.35$ and for different exchange coupling $m_0$ at fixed $\xi_m=1$. The distance between the magnetic and the superconducting vortex is normalized with respect to $\xi_m$. Here, $\Delta=0.3$, $\mu=0$, $t=1$.}
    \label{fig:fig2}
\end{figure*}

\subsection{Microscopic model}\label{sec:model}
We investigate the effect of coupling magnetic and superconducting vortices present in a two-dimensional heterostructure comprising a ferromagnet and an $s-$wave superconductor. Our microscopic model focuses on the superconductor, incorporating the effect of the underlying magnetic layer through an exchange coupling between the electron spins and the magnetic moment. The spin configuration of the magnet is considered to be frozen in this situation. 

The Hamiltonian for an $s$-wave superconductor with spin-orbit coupling in a 2D geometry can be written as $H=\int d^{2}\mathbf{r}\Psi^{\dag}(\mathbf{r})\mathcal{H}(\mathbf{r})\Psi(\mathbf{r})$ where
\begin{eqnarray}\label{eq:hpos}
    \mathcal{H}(\mathbf{r})&=&\left[-\frac{\nabla^{2}}{2m^{*}}-\mu+i\alpha(\sigma^{y}\partial_{x}-\sigma^{x}\partial_{y})\right]\tau^{z}\\ \nonumber
    &+&\mathbf{m(r)}\cdot\bm{\sigma}+\tau^{x}{\rm Re}\Delta(\mathbf{r})-\tau^{y}{\rm Im}\Delta(\mathbf{r}),
\end{eqnarray}
with $\Psi^{\dag}(\mathbf{r})=[\psi^{\dag}_{\uparrow}(\mathbf{r}),\psi^{\dag}_{\downarrow}(\mathbf{r}),\psi_{\downarrow}(\mathbf{r}),-\psi_{\uparrow}(\mathbf{r})]$  the Nambu spinor. Here, $\mu$ is the chemical potential, $m^{*}$ denotes the effective electron mass, $\alpha$ is the Rashba spin orbit coupling and $\mathbf{m(r)}$ is the exchange field due to the ferromagnet. $\sigma^\alpha$ and $\tau^\alpha$ are Pauli matrices acting in spin and Nambu space, respectively.

A superconducting vortex with its center at $\mathbf{r}_{s} = (x_{s},y_{s})$  is modeled with open boundary conditions using the  order parameter 
\begin{equation}\label{eq:sc_vortex}
    \Delta(\mathbf{r})=\Delta_{0}\textrm{tanh}\left(\frac{\vert \mathbf{r}-\mathbf{r_{s}}\vert}{2\xi_{s}}\right)e^{i n_s\theta_s},
\end{equation}
where $\vert \mathbf{r}-\mathbf{r_{s}}\vert$ is the distance from the center of the vortex and $\theta_s=\textrm{tan}^{-1}\left(\frac{ y-y_{s}}{x-x_{s}}\right)$ is the phase winding around the vortex core.  In the presence of the SC vortex, we can write $\bm{\nabla}\theta_s = n_s \hat{\phi}_s/r_s$ where $(\phi_s,r_s)$ are polar coordinates in the frame where the superconducting vortex core lies at the origin and $n_s$ is its vorticity.

To this we add the ferromagnet with a vortex centered at  $\mathbf{r}_{m}=(x_{m},y_{m})$ with a magnetization profile parameterized as in Eq.\ \eqref{eq:spin_texture}. The helicity profile is chosen as $\Phi = n_m \phi + \varphi$ with $\phi$ the polar angle, $n_m$ denoting its vorticity and $\varphi$ being the helicity.
In addition we take $\Theta\mathbf{(r)}=\frac{\pi}{2}\tanh{\left(\vert \mathbf{r}-\mathbf{r_{m}}\vert/{2\xi_{m}}\right)}$, with $\xi_m$ controlling the size of the vortex core.   

For numerical calculations it will be convenient to regularize the Hamiltonian Eq.\ \eqref{eq:hpos} on the square lattice leading to the BdG lattice Hamiltonian of the form
\begin{eqnarray}\label{eq:model_pos}
    H_{\text{sc}}&=&\sum_{\br,s,s'}\{c^{\dag}_{\br;s}(4t-\mu)\delta_{ss'} c_{\br;s'}\\ \nonumber
    &+&[c^{\dag}_{\br;s}(-t\delta_{ss'}-iu\sigma^{y}_{ss'})c_{\br+\bf{x};s'} +\text{h.c.}]\\ \nonumber
    &+&[c^{\dag}_{\br;s}(-t\delta_{ss'}+iu\sigma^{x}_{ss'})c_{\br+\bf{y};s'} +\text{h.c.}]\}\\ \nonumber
    &+&\sum_{\br}[\Delta_{\br}c^{\dag}_{\br\uparrow}c^{\dag}_{\br;\downarrow}+\text{h.c.}]
\end{eqnarray}
Here $\br$ labels the coordinates of lattice sites, $t$ is the nearest neighbor hopping amplitude, $\Delta_{\br}$ is the superconducting pair field, and $u$ is proportional to the spin-orbit coupling parameter, $\alpha$. The exchange field of the ferromagnet is included as follows
\begin{equation}
    H_{\text{m}}=\sum_{\br,s,s^{'}}c_{\br,s}^{\dag}\left[m(r)_{\alpha}\sigma^{\alpha}_{ss'}\right]c_{\br,s^{'}},
\end{equation}
where $\mathbf{m(r)}$ contains the spin texture of the ferromagnet. The Hamiltonian of the composite system is then 
\begin{equation} \label{eq:compositeH}
    H=H_{\text{sc}}+H_{\text{m}}.
\end{equation}

In the next part, we analytically study the interaction between a magnetic and superconducting vortex using an effective magnetoelectric free-energy density. We then compare our analytical results with full numerical simulations of the free-energy density using the composite Hamiltonian, Eq.~(\ref{eq:compositeH}). We find that in certain ranges of the SOC strength ($\alpha$) and the length of the magnetization ($m_0$) we get a stable structure for a superconducting vortex bound to a magnetic vortex.

\subsection{Magnetoelectric free-energy}\label{sec:free_energy}

Our heterostructure explicitly breaks inversion symmetry which allows a Rashba SOC term in the superconductor. This SOC induces a corresponding magenetoelectric interaction between the supercurrent induced by the spin moments of the ferromagnet and the supercurrent density already present in the superconductor, say from a vortex. Note that this current-current interaction can be obtained by considering an exchange interaction between the spin moments of each system and then integrating out the fermions through their propagator \cite{schaffer-magnetoelectric-sc}. 

The coupling can be expanded in powers of the SOC strength $\alpha$\cite{schaffer-magnetoelectric-sc}. We now carry out the calculation of this energy density for a magnetic vortex interfaced with a superconducting vortex in order to study the stability of the bound state between the two defects. The linear coupling between the magnetization $\mathbf{m}(\mathbf{r}) = m_0\mathbf{\hat{m}}(\mathbf{r})$, see Eq.~\eqref{eq:spin_texture}, and the supercurrent $\mathbf{J}_{s} \propto \bm{\nabla}\theta_s/2+{\bf A}$ is
\begin{equation}\label{eq:first-order-ME}
    F_{ME} = \kappa \int d\mathbf{r} (\mathbf{\hat{z}}\times\mathbf{m})\cdot\left(\frac{\bm{\nabla}\theta_s}{2} + \mathbf{A}\right),
\end{equation}
where $\mathbf{A}$ is the magnetic vector potential and $\kappa \propto \alpha$.

In the heterostructure we consider, we place a magnetic vortex at the origin. The profile of the magnetic vortex is given by $\mathbf{m}(\mathbf{r})$ with the boundary conditions on the $\Theta$ field chosen as $\Theta(r = 0) = 0$ and $\Theta(r\rightarrow\infty) = \pi/2$. In addition, we have the $\Phi$ field vorticity, $n_m$, for a contour around the core. In the superconducting layer we assume an order parameter with a vortex core which can be displaced from the origin, $\theta_s = n_s \arctan[(y-y_s)/(x-x_s)]$. Here $(x_s,y_s)$ is the location of the superconducting vortex core. This can be used to calculate the gradient of the phase in the planar polar coordinates  
\begin{eqnarray}\label{eq.nabla_theta}
    \frac{\theta_s}{n_s} &=& \arctan\left( \frac{r\sin\phi-r_{s}\sin\phi_{s}}{r\cos\phi - r_{s}\cos\phi_{s}}\right), \\ \nonumber
    \bm{\nabla}\left(\frac{\theta_s}{n_s}\right) &=& \frac{(r - r_s \cos\Tilde{\phi})\bm{\hat{\phi}}- 2 r r_s \cos {\Tilde{\phi}} \ \mathbf{\hat{r}}}{r^2 + r_s^2 - 2 r r_s \cos \Tilde{\phi}},
\end{eqnarray}
where $\Tilde{\phi} = \phi - \phi_s$. In the limit where the penetration depth exceeds the core sizes $\lambda\gg \xi_m,\xi_s$, we can ignore the screening currents $\mathbf{j} = -\mathbf{A}/4\pi\lambda^2$ induced by orbital or dipolar magnetic fields \cite{gubin2005}. Hence, we set $\mathbf{A}=0$. This is similar to the assumption made by \textcite{hals-skyrmion-vortex}. For the magnetic vortex we set the azimuthal field to $\Phi(r,\phi) = n_m \phi + \varphi$, and consider the situation where $n_m = 1$. The resulting free energy is given by
\begin{eqnarray}\label{eq.attraction}
    F_{ME} &=& \kappa n_s m_0 f(r_{sm})\cos\varphi,\\ \nonumber
    f(r_{sm}) &=& \pi\int_{r_{sm}}^{\infty}dr \sin \Theta(r),
\end{eqnarray}
where $r_{sm}$ (in our setup $r_{sm} = r_s$) is the separation between superconducting and magnetic vortex cores. Crucially since we have a magnetic vortex where the $\Theta$ field varies between $[0,\pi/2]$ this integral is always positive definite. Thus the nature of the interaction, attractive or repulsive, between the two cores is controlled by the product $\kappa n_s m_0 \cos{\varphi}$.

Note that this term vanishes for $\varphi = \pi/2$ and is maximum for $\varphi = 0$. This can be anticipated from the supercurrent mediated interaction picture presented in \textcite{schaffer-magnetoelectric-sc}. For $\varphi = 0$, the supercurrents induced by the magnetic vortex are \textit{collinear} with the superconducting vortex supercurrents. This maximizes the interaction between the two. In the case of $\varphi = \pi/2$, the induced supercurrent density from the magnetic vortex is zero to first order in SOC, with some small second order corrections, and hence we have negligible interaction. The transition between the two scenarios is shown in Fig.~\ref{fig:fig5} and discussed in Sec.~\ref{sec:mzm} in further detail. On evaluating the integral in Eq.~(\ref{eq.attraction}), for our chosen magnetization field, we find the interaction energy between the vortex pair is quadratic in the core separation $r_{sm}$, $F_{ME} \propto \xi_m(r_{sm}/\xi_m)^2$.

We can numerically calculate the free energy of the system using the tight-binding Hamiltonian Eq.~\eqref{eq:model_pos}. The free energy, $F$, of the BdG system is evaluated from the energy eigenvalues $E_n$ using the standard formula
\begin{equation}
F=-2k_{\text{B}}T{\sum_n}'\text{ln}\left[2\text{cosh}\left(\frac{E_{n}}{2k_{\text{B}}T}\right)\right],
\end{equation}
where the prime indicates summation over positive eigenvalues $E_n>0$.
In the limit of zero temperature, the above equation simplifies to $F_{T\rightarrow 0}=\sum_n'(-E_{n})$. The formation of the hybrid vortex is energetically favorable if the free energy minimum occurs when the superconducting and the magnetic vortex overlap, i.e. $r_{sm} \to 0$. The conditions for the formation of the hybrid vortex are summarized in Fig.~\ref{fig:fig2}. In these simulations, we fix the vorticity and the helicity of the magnetic vortex as $n_m=1$ and $\varphi=0$ respectively. 

Taking $F_0$ as the free energy when $r_{sm}=0$, we plot $\tilde{F}=F-F_0$ of the hybrid vortex as a function of distance between the magnetic and superconducting vortices in the region of stability for different parameter choices in Fig.~\ref{fig:fig2}. Matching our analytical prediction we see a quadratic dependence of the mutual interaction energy on the core separation, $r_{sm}$. Further, we observe that the minimum of $\tilde{F}$ is stable at $r_{sm}=0$ when the decay length of the magnetic vortex, $\xi_m$, is increased but the parabola becomes narrower indicating that the coefficient of the quadratic interaction is dependent on $\xi_m$. Further, we observe that $\tilde{F}$ remains nearly the same when the magnitude of the exchange coupling, $m_0$, is increased with the minimum occurring at $r_{sm}=0$ as expected. 

We have also numerically verified the linear dependence of $F_{ME}$ on $\alpha$ and $m_0$. We observed that the variation with $\varphi$ follows the $\text{cosine}$ form near $\varphi =  0$ and $\varphi = \pi/2$ but deviates at other helicities due to the contribution of higher order terms.

The second order in SOC term can be written down as
\begin{equation}
    F_{ME}^{(2)} = \beta \int d\mathbf{r} (\bm{\nabla}m_z \times \mathbf{\hat{z}})\cdot\left(\frac{\bm{\nabla}\theta_s}{2} + \mathbf{A}\right),
\end{equation}
with $\beta \propto \alpha^2$. We can calculate this term in the same configuration with the magnetic vortex at the origin and the superconducting vortex at $\mathbf{r}_s$. We will take the limit $\mathbf{r}_{sm}( = \mathbf{r}_s)\to 0$ and look at the leading order in $\mathbf{r}_{sm}$ correction. Setting $\mathbf{A} = 0$ as before the integral reduces to:
\begin{multline}
    F_{ME}^{(2)} =\\
    \frac{ \beta m_0}{2} \int d\mathbf{r}\sin\Theta(r)\left(\frac{\partial\Theta}{\partial r}\mathbf{\hat{r}}\times\mathbf{\hat{z}} + \frac{1}{r}\frac{\partial \Theta}{\partial \phi} \bm{\hat{\phi}}\times\mathbf{\hat{z}}\right) \cdot\bm{\nabla}\theta_s, 
\end{multline}
where $\bm{\nabla}\theta_s$ is shown in Eq.~(\ref{eq.nabla_theta}). We can drop the second term $\partial \Theta /\partial \phi$ when the magnetic vortex is at the origin. The remaining integral can be evaluated after noting that $\mathbf{\hat{r}}\times\mathbf{\hat{z}} = - \bm{\hat{\phi}}$, obtaining 
\begin{equation}  \label{eq;second-order-ME}
     2F_{ME}^{(2)} = \pi \beta m_0 n_s \left(\cos\left[\frac{\pi}{2}\textrm{tanh}\left(\frac{r_{sm}}{2\xi_{m}}\right)\right] - 1\right),
\end{equation}
for $r_{sm} \leq \xi_m$  and  $ 2F_{ME}^{(2)}=\pi \beta m_0 n_s$ for  $r_{sm} > \xi_m$.
To evaluate the exact forms we have used the magnetization profile shown in  Eq.~(\ref{eq:spin_texture}). 

The second order term is bound from above and can take a maximum value of $(\pi/2)\beta m_0 n_s$ unlike the first order term. This in addition to the fact that it appears at a higher order in SOC ensures that the first order term dominates except when $\varphi \to \pi/2$. Since $ F_{ME}^{(2)}$ is independent of the helicity $\varphi$ of the magnetic vortex it dominates when $\varphi \to \pi/2$ where $F_{ME}$ vanishes (see \eqref{eq:first-order-ME}). Even in this situation this term is not sufficiently large enough to produce a stable bound state between the two vortices according to our numeric simulations of the free-energy.

\subsection{Motion of the bound state}
One of the key engineering features of binding superconducting vortices carrying an MZM to the magnetic vortex, is that we can now propel the magnetic vortex in its own layer and hence propel the MZM. One has to take care that the vortices are still bound to each other during the process, and this can be ensured that the inter-core distance, $r_{sm}$ does not exceed the effective radius of the magnetic vortex, controlled by $\xi_{m}$. Within these bounds we use the effective interaction between the two cores
\begin{equation}
    \mathcal{U}_{int} = \frac{k}{2} (\mathbf{r}_m - \mathbf{r}_s)^2,
\end{equation}
where one can extract the value of the spring constant $k$ from our tight binding model. As mentioned above it is dependent on the decay length of the magnetic vortex but is independent of the exchange coupling parameter. 

For our case we analyze the effect of a spin current in the magnetic layer, following \textcite{hals-skyrmion-vortex}. The equations of motion and the steady state velocity are similar for us, with the Skyrmion gyroscopic term replaced by that of the vortex. Notably, the magnetic vortex does not have an inertial mass unlike the magnetic Skyrmion and obeys the Thiele equations \cite{thiele73,Huber1982}. Let us first look at the system in the absence of any damping:
\begin{eqnarray}
    \label{eq.eom-bound-state}
    \mathbf{G}_m \times [\Dot{\mathbf{r}}_m - \mathbf{v}] &=& -k(\mathbf{r}_m - \mathbf{r}_s) \\ \nonumber
    m_s \Ddot{\mathbf{r}}_s + \mathbf{G}_s \times\Dot{\mathbf{r}}_s &=& - k(\mathbf{r}_s - \mathbf{r}_m),
\end{eqnarray}
where $\mathbf{G}_m = 2\pi S n_s p \mathbf{\hat{z}}$ is the gyroscopic tensor for the magnetic vortex, with $S$ as the spin density and $p$ as the polarization, $\text{sgn}(m_z)$, of the magnetic core. The superconducting cortex gyroscopic term, $\mathbf{G}_s = 2\pi \rho_s n_s \mathbf{\hat{z}}$, with $\rho_s$ as the superfluid density. The spin current is an external spin polarized current adiabatically coupled to the ferromagnetic layer \cite{chauleau2010}. It enters as a correction to the time derivative in the equation of motion of the magnetic vortex, and is represented by a velocity $\mathbf{v}$ \cite{dasgupta-conserve}. In the case of a standard magnetic insulator like permalloy $|\mathbf{v}| = \hbar P j/(2 e S)$, where $j$ is the external current density. 

In the steady state limit, $\Ddot{\mathbf{r}}_s = 0$, with $\Dot{\mathbf{r}}_s = \Dot{\mathbf{r}}_m = \Dot{\mathbf{r}}$, i.e. the cores move together.
We obtain
\begin{equation}
    (\mathbf{G}_s + \mathbf{G}_m)\times \Dot{\mathbf{r}} = \mathbf{G}_m\times\mathbf{v},
\end{equation}
leading to a steady state velocity of $|\mathbf{\Dot{r}}| = |\mathbf{v}| [G_m /(G_s + G_m)]$. In the presence of Gilbert damping and field like damping from the spin current the steady state velocity is modified. Going back to the original equations of motion, Eq.~(\ref{eq.eom-bound-state}), in steady state, without the assumption $\Dot{\mathbf{r}}_s = \Dot{\mathbf{r}}_m $, we can see that a situation with $G_m = -G_s$ would result in a steadily growing core separation. This, however, requires fine tuning the systems and is not generic. The strong attraction between the two cores also creates a situation where a drifting superconducting vortex can bind and carry a magnetic vortex, providing long range dissipation-free spin transport. This schematic has been sought after in spintronics and other proposals with magnetic vortices carrying spin currents have been proposed \textcite{spin-transport-yaroslav}.

\section{Majorana modes}\label{sec:mzm}

Majorana bound states (zero modes) are expected to be localized at the cores of vortices in two dimensional superconductors which have a significant spin orbit coupling to mimic  spinless fermions and are acted upon by a Zeeman-like field to break the time reversal symmetry \cite{Sau2010,lutchyn-sau-sds-mzm-wire,yuval-refael-oppen-mzm-wire}. This Zeeman field is essential for accessing the topological region for a pure two-dimensional superconductor. However, for a hybrid vortex, the proximitized spin texture breaks the time-reversal symmetry without an external Zeeman field and induces a manner of spin polarization. We show here that the system with the hybrid vortex hosts Majorana bound states. One of these states is localized at the vortex core while the other is at the edge of the disk-shaped topological region centered at the vortex. As we shall explain, the non-trivial topology in a system with a hybrid vortex originates from a novel \textit{orbital} effect mediated by an emergent  magnetic vector potential induced by the magnetic vortex. 

The Majorana modes can be observed as zero-bias peaks in the local density of states (LDOS) defined as 
\begin{equation}
    \rho_{\br}(E)={\sum_{\sigma,n}}'\left[\vert u_{\br\sigma}^{n}\vert^{2}\delta(E_{n}-E)+\vert v_{\br\sigma}^{n}\vert^{2}\delta(E_{n}+E)\right].
\end{equation}
Here the eigenvector at energy $E_{n}$ of the BdG Hamiltonian assumes the form $\psi_{\br,n}=(u_{\br\uparrow},u_{\br\downarrow},v_{\br\downarrow}^{*},-v_{\br\uparrow}^{*})_{n}^\text{T}$ (see Appendix~\ref{sec:app_ldos}).
Note that the spin polarized LDOS can be obtained using the same equation without the summation over the spin degrees of freedom, $\sigma$.

\subsection{Pure superconducting and magnetic vortex}

\begin{figure*}
    \centering
    \includegraphics[width=\linewidth]{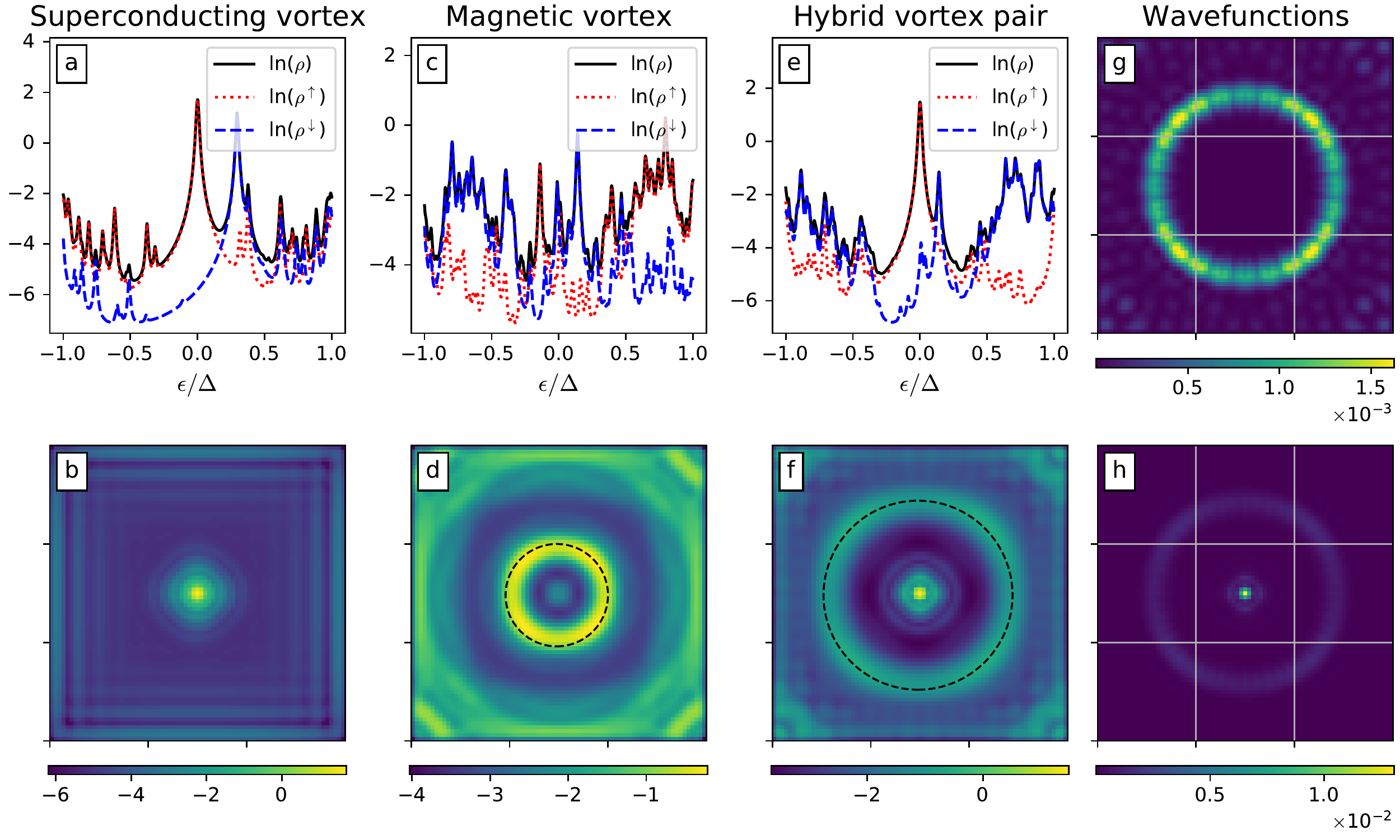}
    
    \caption{\textbf{Panels a-b:} System with only superconducting vortex with external Zeeman field (equivalent to hybrid vortex with $\xi_m\rightarrow\infty)$). (a) The local density of states (LDOS), $\rho(r_{s},\epsilon)$ shows prominent zero-bias peak corresponding to Majorana zero modes. Spin-polarized LDOS reveals that the main contribution to the central Majorana zero-bias peak is from the spin-up LDOS. (b) The 2D plot shows LDOS (log[$\rho(r,\epsilon=0)$]) plotted at zero energy shows a Majorana mode localized at the vortex core. \textbf{Panels c-d:} System with only magnetic vortex (with $\xi_m=1$) without any superconducting vortex. (c) LDOS at the vortex core shows that there is no isolated zero-bias peak. (d) The LDOS plotted at zero energy shows a ring-like structure ($R_{\text{mv}}$) surrounding the vortex core. The black dotted circle depicts the radius at which the local excitation gap vanishes (see Eq.~\eqref{eq:gap_mv}). \textbf{Panels e-h:} System with hybrid vortex ($\xi_m=\xi_s=1$). (e) The local density of states (LDOS), $\rho(r_{s},\epsilon)$ shows prominent zero-bias peak corresponding to Majorana zero modes. Spin-polarized LDOS reveals that the main contribution to the central Majorana zero-bias peak is from the spin-up LDOS as expected when $n_s=n_m$. (f) The 2D plot shows LDOS (log[$\rho(r,\epsilon=0)$]) plotted at zero energy shows a Majorana mode localized at the vortex core. Additionally, there is a ring (radius $R_{\text{hv}}$) surrounding the core where the excitation gap vanishes. The radius shown by the dotted black circle, is estimated where the local excitation gap in equation Eq.~\eqref{eq:gap_hvp_app} vanishes. (g-h) The decoupled wavefunction probabilities, $|\phi_{e}|^{2}$ and $|\phi_{c}|^{2}$, corresponding to the two zero modes show that one Majorana mode is localized at the core while the other is present at the outer ring surrounding the vortex. Parameters: $L=75$, $\Delta=0.3$, $m_{0}=-0.35$, $\varphi=0$, $u=0.6$, $\mu=0$, $t=1$.}
    \label{fig:fig3}
\end{figure*}

Before we study the Majorana modes in hybrid vortices, let us look at the systems where only one of the vortices is present. When only superconducting vortex is present, a Zeeman field is required to create a topological phase. This setup is equivalent to the system given by the BdG Hamiltonian in Eq.~\ref{eq:hpos} in the limit where the magnetic vortex decay length is $\xi_m\rightarrow\infty$ which implies $\mathbf{m}=\mathbf{m_z}=m_0 \mathbf{\hat{z}}$.

In this case, the system is in a topological phase when
\begin{equation}\label{crit}
 m_{z}^{2}>\Delta_{0}^{2}+\mu^{2}
\end{equation}
 and Majorana zero modes are present at the vortex core and at the outer edge of the sample. The local density of states for this system shown in panels (a) and (b) of Fig.~\ref{fig:fig3} reveals a prominent zero bias peak localized at the vortex core. The spin-polarization in the LDOS is expected due to the time-reversal breaking Zeeman field provided by $m_z$ in this case. 

Now, let us look the system with a magnetic vortex and uniform superconducting order parameter. In this situation we observe a high density of states near zero energy in the spectrum. This is evident in  Fig.~\ref{fig:fig3}\textcolor{blue}{c} where the distinct Majorana zero-bias peak in the LDOS is absent but a large number of low lying excitations dominate the spectrum. Fig.~\ref{fig:fig3}\textcolor{blue}{d} indicates that these are concentrated along a ring-like structure surrounding the magnetic vortex. The presence of low energy excitations suggests that the excitation gap vanishes at the ring, as would be the case if this were a boundary between a topological and a trivial phase. However, because $\xi_m=1$ in this plot the magnetization is entirely in plane long before one reaches the ring and the condition Eq.\ \eqref{crit} cannot explain the apparent topological phase inside the ring. As we argue below, the topological phase inside the ring instead owes its existence to a novel orbital effect associated with the spatially varying magnetization field in the vicinity of the vortex center.

As explained in Refs.\ \cite{lutchyn-sau-sds-mzm-wire,yuval-refael-oppen-mzm-wire} the existence of the topological phase in the general class of models considered here depends on the nature of the excitation gap at the origin of the momentum space ${\bf p}=0$. We thus examine the low energy effective Hamiltonian obtained by expanding Eq.\ \eqref{eq:hpos} to leading order in small momenta,
\begin{equation}\label{dirac}
    H_{\rm eff}=\alpha(\bm{\sigma}\times \mathbf{p})_z\tau_z+\Delta_0\tau_x+(\mathbf{m}\cdot\bm{\sigma})\tau_0,
\end{equation}
where $\mathbf{m}=(m_x,m_y,m_z)$. This can be rearranged into a more revealing form
\begin{multline}\label{eq:orb_mag}
H_{\rm eff}=\\
\begin{pmatrix}
\alpha[\bm{\sigma}\times(\mathbf{p}-\mathcal{A})]_z & \Delta_0\\
\Delta_0 & -\alpha[\bm{\sigma}\times(\mathbf{p}+\mathcal{A})]_z
\end{pmatrix}
+m_z\sigma_z,
\end{multline}
where we have identified the in-plane magnetization with an emergent gauge potential $\mathcal{A}=\left(\frac{m_y}{\alpha},-\frac{m_x}{\alpha}\right)$ minimally coupled to Dirac electrons. The system therefore is subjected to a magnetic field given by $\mathcal{B}=\bm{\nabla}\times\mathcal{A}$. This field can be calculated explicitly, away from the vortex core using the magnetic texture $\mathbf{m} = m_0(\cos\Phi,\sin\Phi)$ with $\Phi = \phi + \varphi$. Here the core is placed at the origin and $\varphi$ is the helicity discussed in Sec.~\ref{sec:free_energy}. This gives 
\begin{equation}\label{eq:induced mag}
    \mathcal{B} =  \frac{n_m m_0}{r}\cos\varphi~ \mathbf{\hat{z}}. 
\end{equation}
It is well known that application of a uniform magnetic field $B$ to Dirac electrons in 2D rearranges their spectrum into a set of discrete Landau levels, thus creating an excitation gap $\propto\sqrt{B}$. Here, according to Eq.\ \eqref{eq:induced mag}, we are dealing with a non-uniform magnetic field  that decays as $1/r$ away from the vortex center. Nevertheless, in a semiclassical approximation we might expect the emergent field $\mathcal{B}$ to produce an excitation gap $\propto 1/\sqrt{r}$ locally. Then, in analogy with Eq.\ \eqref{crit} the ring observed 
in Fig.~\ref{fig:fig3}\textcolor{blue}{d}, can be interpreted as marking the edge of the topological region, where the magnetic gap dominates over the SC gap.

To find the radius of the ring, we estimate in Appendix~\ref{sec:app_egap} the size of the energy gap produced by the emergent magnetic field away from the vortex core. This local excitation gap, $E_{g}$, to lowest order in spin field gradients reads
\begin{equation}\label{eq:gap_mv}
    E_g^{2}=\Delta_0^{2}+m_0^2- 2\sqrt{\Delta_0^2m_0^2+u^2\mathcal{B}^2}.
\end{equation}
The excitation gap vanishes at a radius $R_{\text{mv}}$ marking the topological region. This coincides with the radius of the ring in the LDOS simulation shown in Fig.~\ref{fig:fig3}\textcolor{blue}{d}. In the subsequent discussion, it will be clear that these ring-like features are essential in the characterization of the topological regions for the hybrid vortex.

\subsection{Hybrid vortex}\label{sec:hvp}

\begin{figure*}
    \centering
    \includegraphics[width=\linewidth]{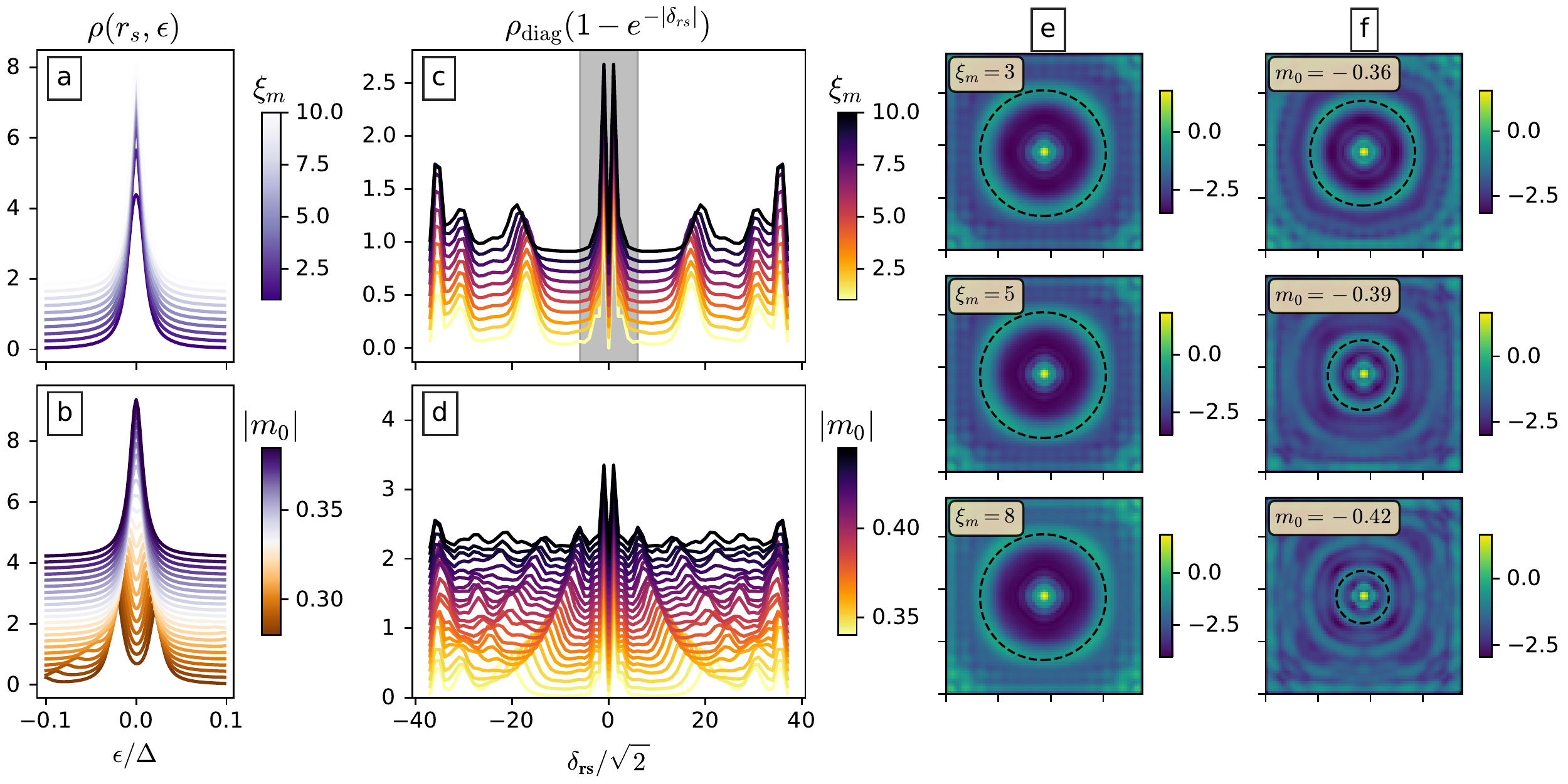}
    \caption{Majorana modes in system with a hybrid vortex for different parameters. \textbf{(a)} The local density of states (LDOS), $\rho_{r_s,\epsilon}$ shows prominent zero-bias peak corresponding to Majorana zero modes for varying magnetic vortex decay lengths, $\xi_m$, for fixed exchange coupling strength, $m_0=-0.35$. \textbf{(b)} LDOS at the vortex core for varying exchange coupling, $m_0$, at fixed magnetic vortex decay length, $\xi_m=1$. LDOS exhibits a zero-bias peak for $|m_0|\gtrsim0.33$. Parameters: $L=75$, $\Delta=0.3$, $\xi_s=1$, $u=0.6$, $\mu=0$, $t=1$. \textbf{(c)} LDOS in hybrid vortex at different values of magnetic vortex decay length, $\xi_m$ (with $m_0=-0.35$). The plot shows a diagonal cross-section of the LDOS at zero energy for different values of $\xi_m$ (values depicted by the colour bar). The central zero-bias peak suppressed in order to highlight the density at the outer ring using the envelope function $(1-\mathrm{e}^{-|\mathbf{\delta_{rs}}|})$ where $|\mathbf{\delta_{rs}}|=|\br-\br_{s}|$.  Outside the shaded grey region, the spin texture flattens out (up to $10^{-5}~m_{z}^{\mathrm{center}}$). The peaks corresponding to the ring structure lie outside the shaded region remain pinned at the same location with increasing decay length until $\xi_m\leq8$ and begins to diverge thereafter due to edge effects. \textbf{(d)} LDOS in hybrid vortex at different values of exchange coupling $m_0$ (with $\xi_m=1$). The plot shows a diagonal cross-section of the LDOS at zero energy with the central zero-bias peak suppressed. The outer ring decays with increasing $m_0$. \textbf{(e)} Representative 2D zero-energy LDOS plots are shown at magnetic vortex decay lengths, $\xi_m=(3,5,8)$. \textbf{(f)} Representative 2D zero-energy LDOS plots are shown at magnetic vortex decay lengths, $m_0=(-0.36,-0.39,-0.42)$. $R_{\text{hv}}$ calculated from Eq.~\ref{eq:rad} is shown by the dotted black circles in panels (e) and (f). Parameters: $L=75$, $\Delta=0.3$, $\varphi=0$, $u=0.6$, $\mu=0$, $t=1$.}
    \label{fig:fig4}
\end{figure*}

Now, we explore the characteristics of the Majorana zero modes present in the hybrid vortex. We look at the case where the magnetic and superconducting vortex have a similar core size ($\xi_m=\xi_s=1$) and choose $m_0$ such that $m_{z}^2 > \Delta_0^2 + \mu^2$ at the vortex core. In Fig.~\ref{fig:fig3}\textcolor{blue}{e}, we observe a clear Majorana zero-bias peak in the local density of states for the chosen parameters. The contribution to the zero-bias peak is entirely sourced from spin-up sector---reflecting the polarization of the superconductor electrons by the spin vortex. In addition to the localized Majorana mode at the core of the hybrid vortex, we observe a ring surrounding the vortex core  Fig.~\ref{fig:fig3}\textcolor{blue}{f}.
On plotting the decoupled Majorana wavefunctions (see Appendix~\ref{sec:app_wf}) in Fig.~\ref{fig:fig3}\textcolor{blue}{g-h}, we notice that the Majorana modes are localized at the vortex core and at the ring away from the core. 

As in the case of a pure magnetic vortex, we interpret the ring as the edge of the topological region originating from the orbital magnetic field from the in-plane magnetization (see Eq.~\eqref{eq:orb_mag}). We can estimate the radius of the ring by comparing it with the radius at which the local excitation gap vanishes. The local excitation gap for the hybrid vortex includes contribution from the gradients of the phase of both the superconducting and the magnetic vortex which modifies the expression for the local excitation gap as discussed in  Appendix~\ref{sec:app_egap}. We denote $R_{\text{hv}}$ as the radius of the topological region at which this excitation gap vanishes -- shown by the dotted circle in Fig.~\ref{fig:fig3}\textcolor{blue}{f}. The estimated radius again  coincides with the radius of the ring from the LDOS simulations.

Majorana zero modes are present at the core of the hybrid vortex for a range of parameters as demonstrated in panels (a) and (b) of Fig.~\ref{fig:fig4}. Here, we observe a prominent zero-bias peak in the LDOS that is robust for different values of the decay length, $\xi_m$, which controls the size of the magnetic vortex. We also find the critical value of magnitude of the exchange coupling parameter, $m_0$, below which the zero-bias peak splits ($\approx0.33$ for the chosen parameters and magnetic vortex profile). 

\begin{figure*}
    \centering
    \includegraphics[width = \linewidth]{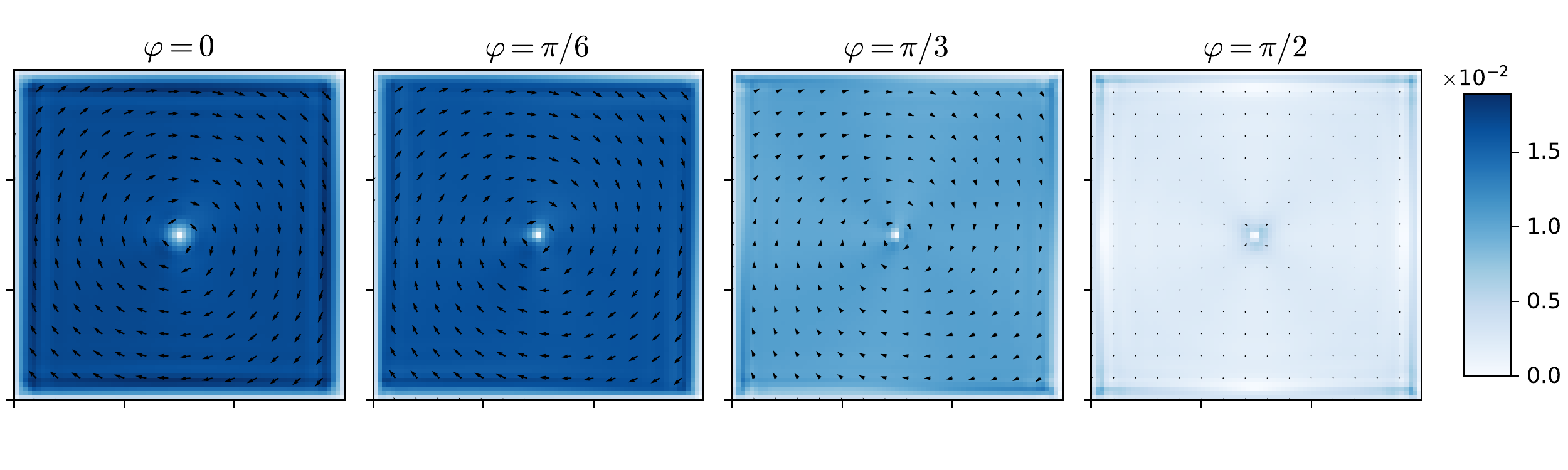}
    \caption{Supercurrents in a system with uniform superconducting order parameter and a magnetic vortex with helicities $\varphi=\left(0,\frac{\pi}{6},\frac{\pi}{3},\frac{\pi}{2}\right)$ (left to right). The phase of the superconducting order parameter is given by Eq.~\eqref{eq:mv_phi_s}. Parameters: $L=75$, $\Delta=0.3$, $m_0=-0.3$, $\xi_s=\xi_m=1$,  $u=0.6$, $\mu=0$, $t=1$.}
    \label{fig:fig5}
\end{figure*}

To study the spatial features of the Majorana modes present in the hybrid vortex we plot the LDOS at zero energy across a diagonal cross-section of the lattice for different decay lengths of the magnetic vortex in panel (c) of Fig.~\ref{fig:fig4}. The value of $R_{\text{hv}}$ is much larger than the radius at which the spin texture flattens out with $m_z$ nearly zero ($m_z \sim 10^{-5}~m_{z}^{\mathrm{core}}$).  Further, we observe that there is no significant difference in the location of the peaks corresponding to the outer ring upon variation of $\xi_m$.  This further supports our interpretation of the topological phase as being enabled by the orbital magnetic effect through the emergent vector potential $\mathcal{A}$ introduced in the previous subsection.
We also note here that at larger values of $\xi_m$, the wavefunction corresponding to outer Majorana mode can exhibit delocalisation in the region ranging from the ring to the edge of the system suggesting the presence of edge effects.

Now, we look at the spatial features of zero-energy LDOS while varying the magnitude of the exchange coupling $m_0$ at fixed value of $\xi_m$ plotted in panel (d) of Fig.~\ref{fig:fig4}. The vortex bound zero-energy Majorana state persists for the entire range of parameters shown in Fig.~\ref{fig:fig4}\textcolor{blue}{d}. The radius of the outer ring, present in the LDOS decreases with the increase in exchange coupling strength. This decrease in the radius of the topological region is expected from the calculated $R_{\text{hv}}$ (see Eq.~\eqref{eq:rad}) as further discussed in Appendix~\ref{sec:app_egap}.

\subsection{Supercurrents}

\begin{figure*}
    \centering
    \includegraphics[width=\linewidth]{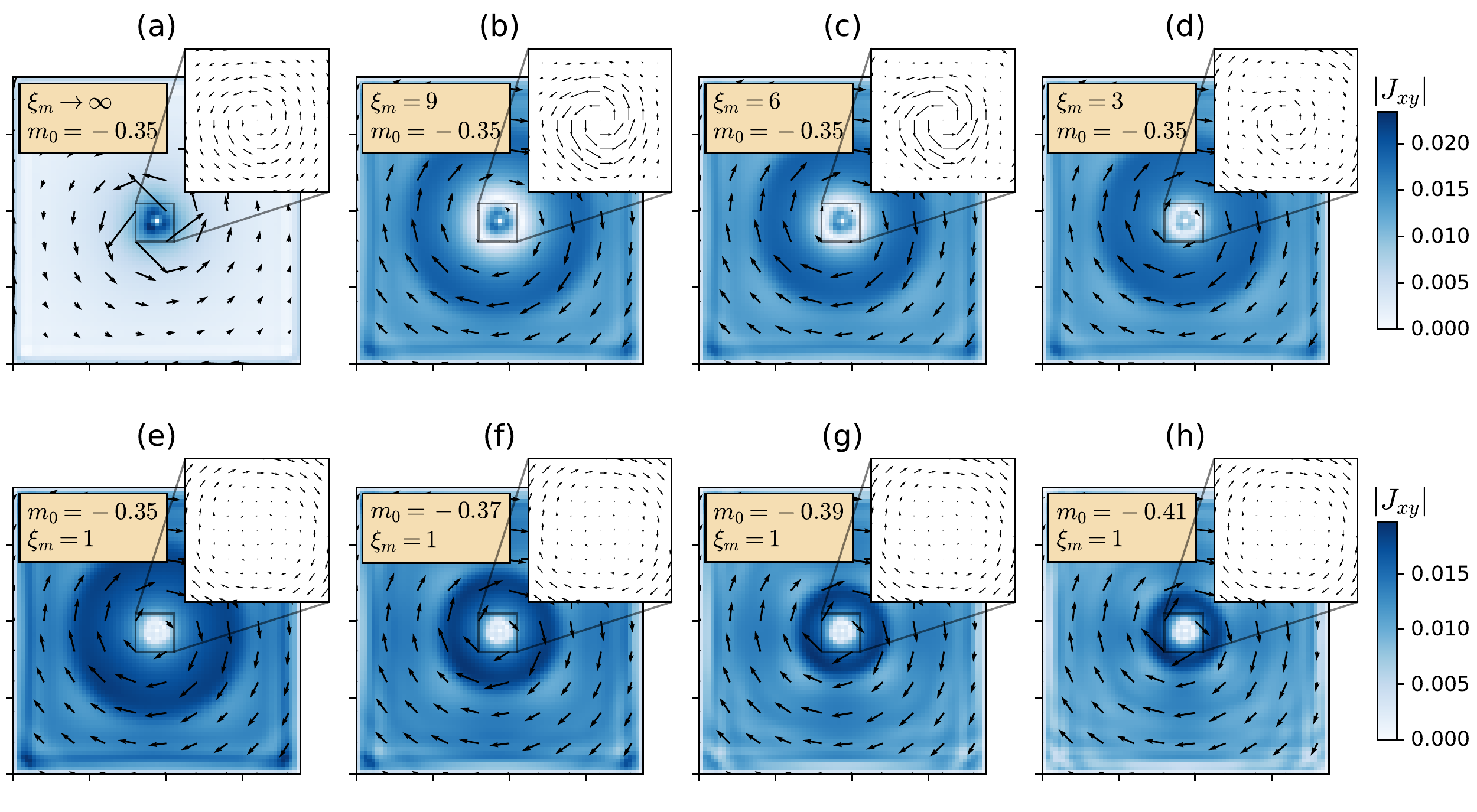}
    \caption{\textbf{(a)-(d)}  Local supercurrents plotted for different values of magnetic vortex decay lengths, $\xi_m=(\infty,9,6,3)$. Case (a) corresponds to a system with superconducting vortex and uniform Zeeman field. \textbf{(e-h)} Local supercurrents plotted for exchange coupling strengths, $m_0=(-0.35,-0.37,-0.39,-0.41)$. The vectors (arrows) correspond to the direction of the flow of the supercurrent given by the vector $\mathbf{J}_{xy}$ . The colorbars represent the magnitude of $J_{xy}$. The insets show the orientation of the local supercurrents around the vortex core. parameters: $L=75$, $\Delta=0.3$, $\xi_s=\xi_m=1$, $\varphi=0$,  $u=0.6$, $\mu=0$, $t=1$.}
    \label{fig:fig6}
\end{figure*}

To understand the characteristic features of hybrid vortices, we study the local supercurrents in these systems. As derived in Appendix \ref{sec:app_ixy} supercurrent flowing along the bond connecting sites $\br$ and ${\br}'$ is given by, 
\begin{equation}\label{eq:local_Isc}
   J_{ij}=-\frac{ie}{\hbar}\sum_{\sigma,\sigma'}\left(\sum_n h_{i\sigma,j\sigma'} u_{i\sigma}^{n}u_{j\sigma'}^{n*}f(E_{n})-\text{c.c.}\right),
\end{equation}
where we denote the matrix element between site $i$ (spin $\sigma$) and site $j$ (spin $\sigma'$) as $h_{i\sigma,j\sigma'}$. We denote the bond current from $x$ to $x+1$ on the lattice as $J_x$ and that from $y$ to $y+1$ as $J_y$. The local supercurrent vector is then given by $\mathbf{J}_{xy}=(J_x,J_y)$. 

To understand the origin of this supercurrent analytically, let us revisit Eq.~\eqref{eq:orb_mag} where we incorporated the in-plane magnetization into the magnetic vector potential $\alpha\mathcal{A}=\left(m_y,-m_x\right)$. The supercurrent density is generally given by 
\begin{equation}\label{eq.london}
    \bm{j}(\bm{r})=n_s\left(\bm{\nabla}\theta_s-2\mathcal{A}\right),
\end{equation}
where $n_s$ denotes the superfluid density. In the self-consistent solution the phase field $\theta_s(\br)$ will adjust so as to minimize the free-energy cost of the current, $\sim ({\bm{\nabla}\theta_s}-2\mathcal{A})^2$. Because $\nabla\times\nabla \theta_s=0$, except for isolated points -- vortices -- only the longitudinal part of $\mathcal{A}$ can be screened completely; the transverse part, corresponding to nonzero field $\mathcal{B}$, can only be screened partially by incorporating vortices into the phase field. More details on this and a relevant calculation are provided in Appendix~\ref{sec:app_supercurrents_mv}.
 We know from Eq.~\eqref{eq:induced mag} that $|\mathcal{B}|\propto \cos\varphi$, implying that supercurrents should decrease as the helicity parameter increases from $0$, eventually vanishing as $\varphi\to\pi/2$ and $\mathcal{A}$ becomes purely longitudinal. This expectation is indeed borne out in the microscopic calculation of the supercurrent displayed in  Fig.~\ref{fig:fig5} for several values of the helicity parameter. An interesting feature of the current flow is its constant magnitude independent of the radius $r$. This is to be contrasted with a regular Abrikosov vortex where the magnitude decays as $|\nabla\theta_s|\sim 1/r$. This peculiar behavior can be traced back to the fact that the current here is proportional to the magnetization field of the magnetic vortex which itself retains a constant magnitude independent of $r$. 

In a hybrid vortex we expect the usual vortex current pattern with the amplitude decaying as $1/r$ to be superimposed on the anomalous current distribution shown in Fig.~\ref{fig:fig5}. For different values of magnetic vortex decay lengths $\xi_m$ this is displayed in Fig.~\ref{fig:fig6}\textcolor{blue}{a-d}. The case where $\xi_m\rightarrow\infty$, shown in Fig.~\ref{fig:fig6}\textcolor{blue}{a}, corresponds to a pure superconducting vortex subject to an external Zeeman field, $m_z$. The supercurrents now flow counter-clockwise and follow the expected $1/r$ behavior outside the vortex core.  
For finite decay lengths, the hybrid vortex exhibits clockwise supercurrents away from the core while closer to the core the currents are counter-clockwise. This can be understood as a competition between the $1/r$ Abrikosov contribution and the $r$-independent magnetization contribution. The former partially cancels the latter at intermediate distances, thus lowering the free-energy cost of the hybrid vortex compared to the pure magnetic vortex. This cancellation also provides some intuition behind the notion of stability of the hybrid vortex.  Another interesting feature to note is the presence of the outer ring like structure in the magnitude of the supercurrents. The outer edge of these supercurrent rings also correspond to the ring in the LDOS observed earlier in Fig.~\ref{fig:fig4}.

We perform a similar analysis for the local supercurrents with varying exchange coupling parameter, $m_0$ and plot the results in Fig.~\ref{fig:fig5}\textcolor{blue}{e-h}. As before, we observe the clockwise supercurrent flow away from the core of the vortex. Closer to the vortex core, we expect the supercurrents to reverse the direction. Note that in the representative plots shown in Fig.~\ref{fig:fig5}\textcolor{blue}{e-h}, the reversal of the supercurrent is not immediately evident owing to the fact that magnetic and the superconducting vortex are approximately of the same size. Here again, we see the presence of a prominent outer ring in the magnitude of the supercurrents. Quite interestingly, the radius of this outer ring decreases with increasing magnitude of the exchange coupling, similar to the outer ring in the LDOS plots Fig.~\ref{fig:fig4}.

\section{Discussion}
We performed a study of a novel 
`hybrid' vortex comprising a superconducting and magnetic vortex. Such a composite can occur in a quasi-2D heterostructure formed by a thin easy-plane ferromagnet and a superconductor with Rashba spin-orbit coupling. The magnetic and superconducting vortices present in this composite structure can form an energetically favorable hybrid vortex. We showed this analytically by considering the magneto-electric interaction induced by the spin-orbit coupling as well as by detailed numerical simulations of a minimal microscopic model.

We established the presence of the zero-energy Majorana bound states in the hybrid vortex. Using the characteristic zero-bias peak in the local density of states, we studied the features of these Majorana modes. In addition to the Majorana zero mode localized at the core of the hybrid vortex, we also observed its partner zero mode localized in the ring-like structure surrounding the hybrid vortex. The ring demarcates a disk-shaped topological region centered at the vortex core. Remarkably, the topological phase inside the ring is stabilized not by an out of plane magnetization $m_z$, as would be the case in conventional models, but by an emergent orbital magnetic field. This emergent orbital magnetic field arises from the magnetization field of the spin vortex which couples as vector potential to electrons in the superconductor.
We derived an estimate for the ring radius using the emergent magnetic field concept and found that it agrees quantitatively with our numerical results. 
We also studied the dependence of the topological ring radius on the helicity of the magnetic vortex which likewise supports the general picture that we presented.

The superflow pattern around the hybrid vortex shows an unusual behavior in that the supercurrent generically reverses its direction at some intermediate radius. This can be understood as a competition between the usual Abrikosov vortex superflow that decays as $1/r$ and an anomalous magnetization-induced current that is $r$-independent. This partial cancellation of the net current  near the reversal region also provides intuition for the energetic stability of the hybrid vortex.

As future avenues of research, the theory for single hybrid vortex can be extended to multiple hybrid vortices. For instance, the energetics in the two hybrid vortices would now need to account for the repulsion between two magnetic vortices in addition to the attraction between the superconducting and magnetic vortices of different pairs. This makes the extension to multiple hybrid vortices an interesting problem. Additionally, a detailed analysis of the Majorana modes with the helicity of the magnetic vortex would be very useful in further understanding the Majorana modes in hybrid vortices. 

\section*{Acknowledgments}
We would like to thank Oguzhan Can and Rafael Haenel for helpful discussions.  This research was supported in part by NSERC and the Canada First Research Excellence Fund, Quantum Materials and Future Technologies Program. S.D. is additionally supported by the Japan Society for the Promotion of Science KAKENHI (Grant No.~JP19H01808).

\bibliography{ref}
\appendix
\renewcommand\thefigure{\thesection.\arabic{figure}} 

\section{Details of the numerics}
\subsection{Local density of states}\label{sec:app_ldos}
Consider the particle number operator on site $i$ given by $N_{i}=\sum_{\sigma}N_{i\sigma}$ with $N_{i\sigma}=c_{i\sigma}^{\dag}c_{i\sigma}$. The eigenvector at energy $E_{n}$ after diagonalization of the BdG Hamiltonian assumes the form $\psi_{i, n}=(u_{i\uparrow},u_{i\downarrow},v^{\dag}_{i\downarrow},-v^{\dag}_{i\uparrow})_{n}^\text{T}$. Using the symmetries of the BdG Hamiltonian, we can also write the number operators as
\begin{equation}\label{ni}
    N_{i\sigma}=\sum_{n}\vert u_{i\sigma}^{n}\vert^{2}f(E_{n})
    =\sum_{n}\vert v_{i\sigma}^{n}\vert^{2}(1-f(E_{n})).
\end{equation}
where $f(E_{n})$ is the Fermi-Dirac distribution function. In the zero T limit, we simply take $f(E_{n})=1$ for $E_{n}<0$ and $0$ otherwise.

The probability distribution of the particle wave-function at site $i$ with energy $E_{n}>0$ is evaluated as $\vert \psi_{\text{p}i}^{n} \vert^{2}=\sum_{\sigma}\vert u_{i\sigma}^{n}\vert^{2}$. 

The local density of states is given by
\begin{equation}
    \rho_{i\sigma}(E)=\sum_{\substack{n;\\E_{n}>0}}\left[u_{i\sigma}^{n}\vert^{2}\delta(E_{n}-E)+\vert v_{i\sigma}^{n}\vert^{2}\delta(E_{n}+E)\right].
\end{equation}

\subsection{Majorana wave-functions}\label{sec:app_wf}
The Majorana modes, if present, will manifest themselves as zero-energy peaks in the local density of states. The two states at zero energy in the spectrum can be assumed to be a linear superposition of the two Majorana wavefunctions localized at the core of the vortex and the edge of the topological region respectively. If we denote the Majorana mode creation operators as $\gamma_c$ and $\gamma_e$, the Hamiltonian for the small overlap of the wavefunctions is uniquely defined by their self-conjugation property  $\gamma=\gamma^{\dag}$ as follows
\begin{equation}
    h_{\textrm{eff}}=i\epsilon_0\gamma_c\gamma_e=\frac{1}{2}\Gamma^{\dag}\sigma^{y}\Gamma,
\end{equation}
where $\Gamma=(\gamma_c,\gamma_e)^{\textrm{T}}$. The above equation implies that the wavefunctions $\phi^{\pm}$ corresponding to the eigenvalues of $\pm\epsilon_0$ are related to the zero mode wavefunctions as $\phi^{\pm}=\frac{1}{\sqrt{2}}\left(\phi_c\pm\phi_e\right)$. Inverting this relation gives us the decoupled Majorana modes localised at the vortex core and edge of the topological region, respectively. 

\subsection{Local supercurrents}\label{sec:app_ixy}
To calculate the local supercurrents or bond currents, we begin with the Heisenberg equation of motion for particle at lattice site $i$
\begin{equation}
    i\hbar\frac{\partial\langle N_{i}\rangle}{\partial t}=\langle[N_{i},H]\rangle,
\end{equation}
where $H$ is the tight-binding Hamiltonian. 
Denoting the matrix element between site $i$ (spin $\sigma$) and site $j$ (spin $\sigma^{'}$) as $h_{i\sigma,j\sigma^{'}}$, we can rewrite the Heisenberg equation of motion as follows
\begin{equation}
    i\hbar\frac{\partial\langle N_{i}\rangle}{\partial t}=\left\langle  \sum_{\substack{j\neq i,\\\sigma,\sigma'}}\left(h_{i\sigma,j\sigma'}c_{i\sigma}^{\dag}c_{j\sigma'}-h_{j\sigma^{'},i\sigma}c_{j\sigma'}^{\dag}c_{i\sigma}\right)\right\rangle.
\end{equation}
Using the above relation, the current operator from site $j$ to site $i$ is,
\begin{equation}
    \hat{J}_{ij}=-\frac{ie}{\hbar}\sum_{\sigma,\sigma'}\left(h_{i\sigma,j\sigma'}c_{i\sigma}^{\dag}c_{j\sigma'}-h_{j\sigma^{'},i\sigma}c_{j\sigma^{'}}^{\dag}c_{i\sigma}\right).
\end{equation}
The average bond current is given by taking the expectation value of the current operator with the ground state eigenvectors (for a superconductor, the ground state eigenvectors are given by summing over all negative energies),
\begin{equation}
\begin{split}
    J_{ij}&=-\frac{ie}{\hbar}\sum_{\sigma,\sigma'}h_{i\sigma,j\sigma^{'}}\\
    &{\sum_{n}}'\left(u_{i\sigma}^{n}u_{j\sigma'}^{n*}f(E_{n})
    +\sigma\sigma^{*}v_{i\sigma}^{n}v_{j\sigma'}^{n*}(1-f(E_{n}))-\text{c.c.}\right).
\end{split}
\end{equation}
Using the symmetries of the BdG Hamiltonian (see Eq. \eqref{ni}), the equation of bond current can be simplified to
\begin{equation}\label{eq:local_Isc1}
   J_{ij}=-\frac{ie}{\hbar}\sum_{\sigma,\sigma'}\left(\sum_{h}h_{i\sigma,j\sigma'}u_{i\sigma}^{n}u_{j\sigma'}^{*}f(E_{n})-\text{c.c.}\right). 
\end{equation}

\section{Topological region of the hybrid vortex system}\label{sec:app_egap}

\begin{figure}
    \centering
    \includegraphics[width=\linewidth]{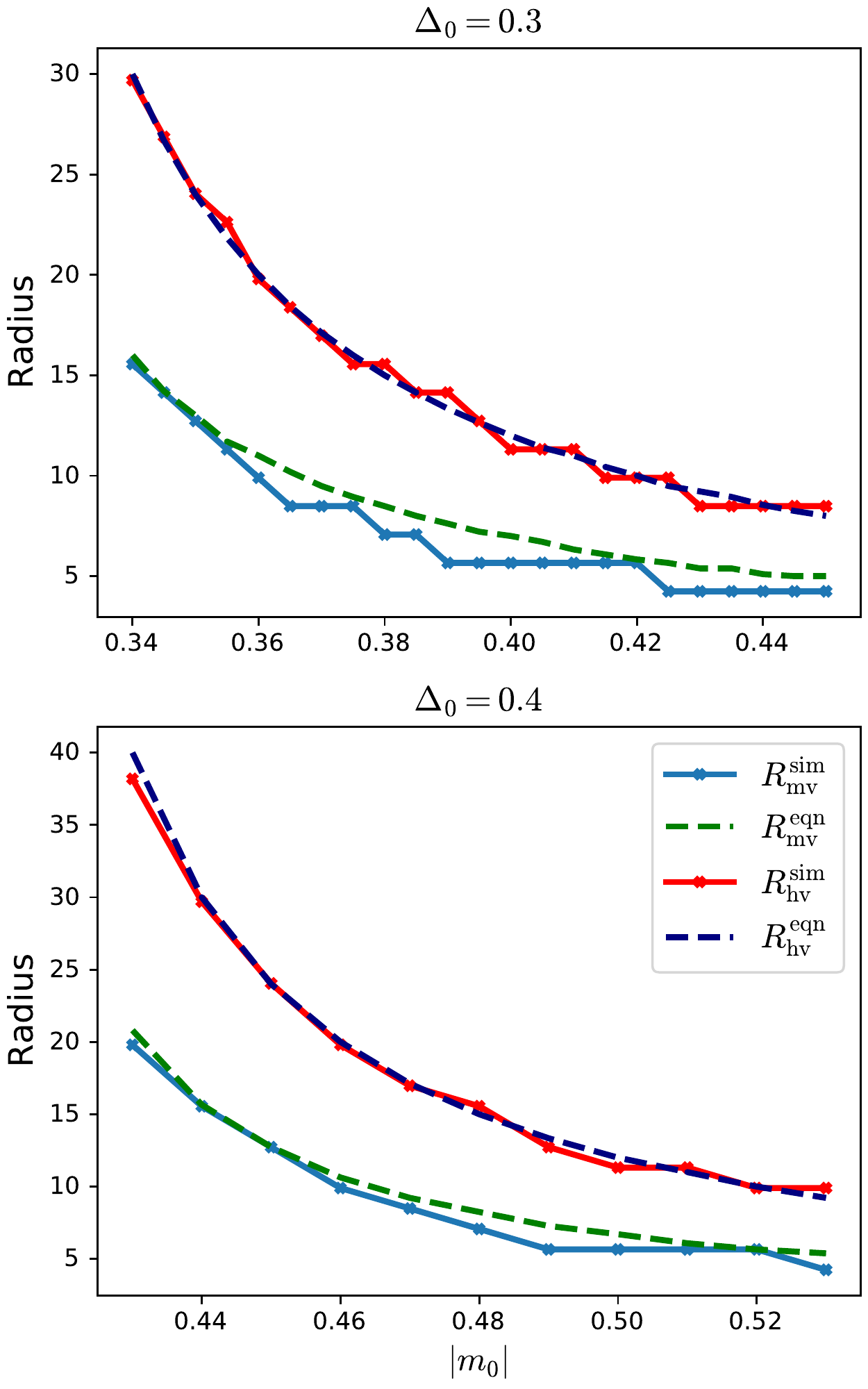}
    \caption{The plot shows the radius of the topological region of the magnetic vortex ($R_{\mathrm{mv}}^{\mathrm{eqn}}$) given by Eq.~\eqref{eq:gap_mv} as a function of the exchange coupling parameter and compares it to the radius of the ring obtained by simulations ($R_{\mathrm{mv}}^{\mathrm{sim}}$) for $\Delta_0=0.3$ (top panel) and $\Delta_0=0.4$ (bottom panel). The plot also compares the radius of the topological region around the hybrid vortex ($R_{\mathrm{hv}}^{\mathrm{eqn}}$) calculated from Eq.~\eqref{eq:gap_hvp_app} and the radius of the ring around the hybrid vortex obtained from the simulations ($R_{\mathrm{hv}}^{\mathrm{sim}}$).}
    \label{fig:f1}
\end{figure}

First, let us look at a pure magnetic vortex with a uniform superconducting order parameter. Away from the vortex core, near zero momenta at $\mu=0$, this is described by the Hamiltonian in Eq.~\eqref{dirac}.
We assume that away from the vortex core, the spins of the ferromagnet are entirely in the $xy$ plane, making the magnetic texture ${\bf m}=m_0(\cos{\phi_m},\sin{\phi_m},0)$. Superfluid velocity in a superconducting vortex is defined as ${\bf v}_s=\bm{\nabla}\theta_s=\frac{1}{r}\hat{\phi}$. Similarly, here we define the magnetic vortex phase gradient as ${\bf v}_m=\bm{\nabla}\phi_m=\frac{1}{r}\hat{\phi}$. Squaring the Hamiltonian in Eq.~\eqref{dirac} and systematically neglecting terms  containing $p^2$ results in 
\begin{equation}
\begin{aligned}
    H^2&=\Delta_0^2+m_0^2+2\Delta_0(\mathbf{m}.\sigma)\tau_x\\
    &+\left\{\alpha(\bm{\sigma}\times\mathbf{p})_z,(\mathbf{m}\cdot\sigma)\right\}\tau_z,
    \end{aligned}
\end{equation}
where the the anti-commutator is defined as $\left\{A,B\right\}=A\cdot B+B\cdot A$. Reshuffling the constant terms and squaring once again using the same assumptions as before, we find
\begin{eqnarray}
    &(H^2-\Delta_0^2-m_0^2)^2 \nonumber\\
    =&4\Delta_0^2m_0^2+\left(\left\{\alpha(\bm{\sigma}\times\mathbf{p})_z,(\mathbf{m}\cdot\sigma)\right\}\right)^2\\
    +&\left\{2\Delta_0(\mathbf{m}\cdot\sigma)\tau_x,\left\{\alpha(\bm{\sigma}\times\mathbf{p})_z,(\mathbf{m}\cdot\sigma)\right\}\tau_z\right\}.   \nonumber
\end{eqnarray}
The anti-commutators in the above equation can be solved by identifying $\mathbf{p}=-i\nabla$ and by considering all spatially dependent functions to be slowly varying such that the second derivatives are negligible,
\begin{eqnarray}
    \left(\left\{\alpha(\bm{\sigma}\times\mathbf{p})_z,(\mathbf{m}\cdot\sigma)\right\}\right)^2=\alpha^2m_0^2v_m^2,
\end{eqnarray}
\begin{eqnarray}
    \left\{2\Delta_0(\mathbf{m}\cdot\sigma)\tau_x,\left\{\alpha(\bm{\sigma}\times\mathbf{p})_z,(\mathbf{m}\cdot\sigma)\right\}\tau_z\right\}=0.
\end{eqnarray}
The resulting energy gap equation is 
\begin{equation}\label{eq:gap_mv_app}
    E_g^{2}=\Delta_0^{2}+m_0^2\pm2\sqrt{\Delta_0^2m_0^2+u^2m_0^2v_m^2},
\end{equation}
where $u=\frac{\alpha}{2a}$ and $v_m^2=\frac{1}{r^2}$ as defined previously. We set the lattice constant $a = 1$. As the minus sign clearly corresponds to a lower energy it will represent the local energy gap. This, together with identifying $m_0^2v_m^2=\mathcal{B}^2$ leads to Eq.\ \eqref{eq:gap_mv} in the main text.

A similar procedure gives us the energy gap for the hybrid vortex where the additional terms arise from the spatial dependence of the phase of the superconducting order parameter. Therefore, for a hybrid vortex, we get the following gap equation
\begin{equation}\label{eq:gap_hvp_app}
    E_g^{2}=\Delta_0^{2}+m_0^2-2\sqrt{\Delta_0^2m_0^2+u^2(m_0\bm{v}_{m}+\Delta_0\bm{v}_{s})^2}.
\end{equation}
where we have again defined $v_s^2=\frac{1}{r^2}$. 

When the gaps given by equations \eqref{eq:gap_mv_app} and \eqref{eq:gap_hvp_app} close, a topological phase transition locally occurs. Setting $E_g=0$ in Eq.~\eqref{eq:gap_hvp_app}, the radius of the topological region for the hybrid vortex is given by
\begin{equation}\label{eq:rad}
    R_{\text{hv}}^{\text{eqn}}=\frac{2u}{\vert |m_0|-\Delta_0 \vert}.
\end{equation}
We plot the radius of the topological region and compare it to the ring around the vortex obtained in the simulations. This is shown in Fig.~\ref{fig:f1}. While Eq.~\eqref{eq:rad} suggests a divergence at $\vert m_0 \vert=\Delta_0$, we note here that the topological region exists only when $|m_0|$ is above a critical value which is always greater than $\Delta_0$. This is also discussed in Sec.~\ref{sec:hvp} (Fig.~\ref{fig:fig4}) where we show the splitting of Majorana zero-bias peak below a critical value, $|m_0|\leq0.33$, for $\Delta=0.3$.

\section{Supercurrents in a pure magnetic vortex}\label{sec:app_supercurrents_mv}
\begin{figure}
    \centering
    \includegraphics[width=\linewidth]{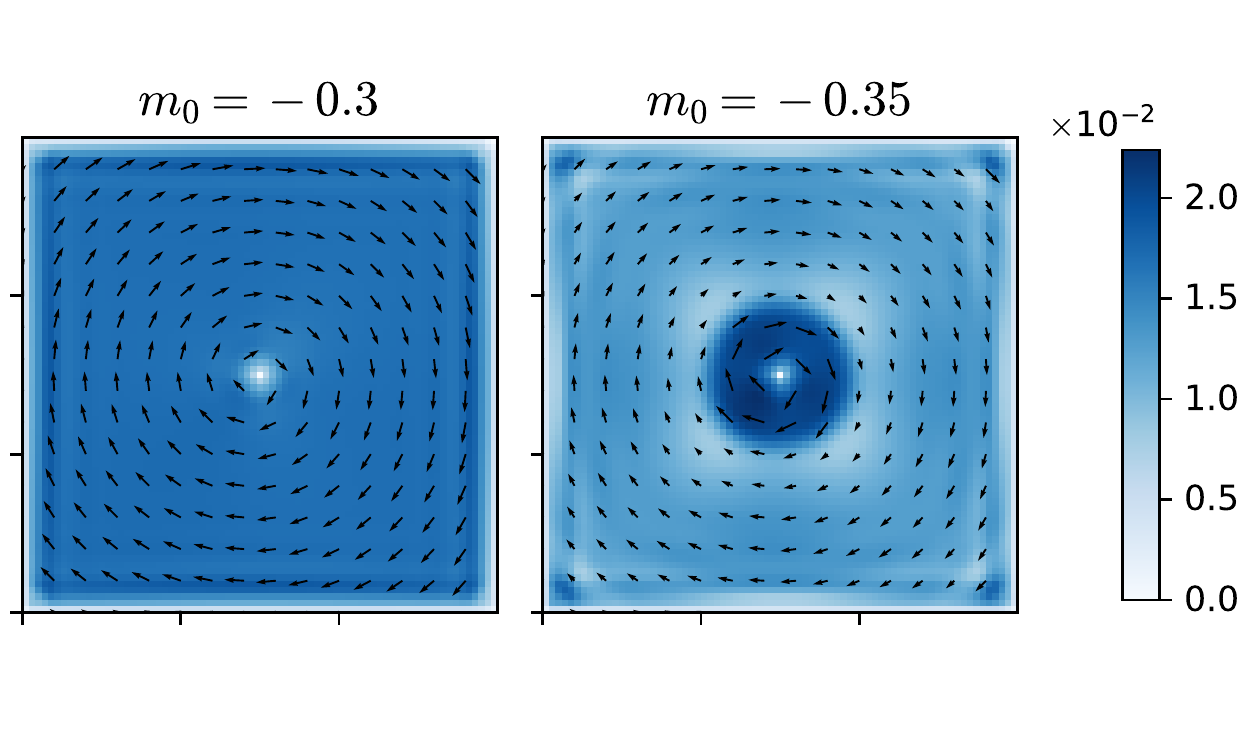}
    \caption{Supercurrents for a pure magnetic vortex in the trivial phase (left) and the topological phase (right). The topological phase transition occurs at the critical exchange coupling $|m_0|\approx0.33$ when $\Delta=0.3$ for the chosen parameters. The phase of the superconducting order parameter is given by Eq.~\eqref{eq:mv_phi_s}.}
    \label{fig:f2}
\end{figure}

To the first order in SOC, the supercurrent is given by\cite{schaffer-magnetoelectric-sc},
\begin{equation}\label{eq.scurrent-an}
    \bm{j}(\bm{r})=\frac{n_s}{2m}\bm{\nabla}\theta_s+\alpha(\hat{z}\times \bm{m}).
\end{equation}
Upon imposing continuity, $\bm{\nabla}\cdot\bm{j}=0$, the solution of the superconducting phase, as derived by \textcite{schaffer-magnetoelectric-sc}, is given by
\begin{equation}\label{eq:mv_phi_s}
    \theta_s=\frac{m\alpha}{n_s}\bm{r}\cdot(\bm{m}\times\hat{z}).
\end{equation}

For our system, $m=1$ and $n_s=1$. To obtain the supercurrents for a pure magnetic vortex shown in Fig.~\ref{fig:fig5}, we include the phase, $\theta_s$, from Eq.~\eqref{eq:mv_phi_s} in the superconducting order parameter. 

We note here that a pure magnetic vortex can by itself induce a topological phase transition in the superconductor since it has a non-coplanar spin ordering \textcite{nagaosa2013}. In our case when the magnitude of the exchange coupling is increased beyond a critical value we see this transition. The signature of the phase transition can be observed in the supercurrents as we tune $m_0$. Above a certain value the supercurrents are no longer uniform throughout the system as expected from \eqref{eq.scurrent-an}, a clear ring forms separating an inner topologically non trivial region including the vortex core. This is shown in Fig.~\ref{fig:f2} for $\varphi=0$.

\section{Modified electromagnetic duality}\label{sec:app_em}
The theory for the planar ferromagnet can be cast into a dual theory of electromagnetism in 2+1 dimension with magnons acting as the single photon and vortices acting as electric charges carrying magnetic fluxes \cite{Kosterlitz1974,dasgupta-2020-vortex}. The duality transformation is between the Lagrangian densities:
\begin{eqnarray}
    S (\cos\Theta - 1) \Dot{\Phi} - \frac{\tilde{A}}{2}(\bm{\nabla}\Phi)^2 - \frac{\mathcal{K}}{2} \cos^2\Theta \\ \nonumber
    \Rightarrow - 2\pi J^\mu \tilde{A}_{\mu} - \frac{F^{\mu\nu}F_{\mu\nu}}{e^2}.
\end{eqnarray}
In the spin Lagrangian $S$ is the spin density $\tilde{A}$ is proportional to Heisenberg exchange and $\mathcal{K}_a$ is the easy plane anisotropy, in the electromagnetic Lagrangian $e^2 = S \tilde{A}/\mathcal{K}_a$. This duality mapping has two parts, the conserved Noether current conjugate to the field $\Phi$ (magnetic spin current $\mathbf{J}_m$) which is mapped through a Bianchi identity to the gauge field $A$ and the electromagnetic tensor. The second part is the matter current $J$ which consists of electrical charges (magnetic vortices \cite{Kosterlitz1974}) which are also conserved due to their topology, and hence follow a continuity equation, for details see \textcite{dasgupta-2020-vortex}. A similar dual construction can be made with the superconducting phase, the conserved current there is the supercurrent, $\mathbf{J}_{s} \propto \bm{\nabla}\theta_s$.

While this duality would have existed in our heterostrucure for an interaction of the form---$\mathbf{J}_m\cdot\mathbf{J}_{s}$, the magnetoelectric interaction does not allow such a construction. This implies that the mapping to electrostatics no longer holds and treating the magnetic vortices and superconducting vortices as interacting charges is not possible.

\end{document}